\documentclass[useAMS,usenatbib,usegraphicx]{mn2e}

\title[The X-ray Luminosity - Velocity Dispersion Relation in the REFLEX Cluster Survey]
  {The X-ray Luminosity - Velocity Dispersion relation in the REFLEX Cluster 
Survey\thanks{Partially based on observations at the European Southern Observatory La Silla, Chile}}
\author[A. Ortiz-Gil et al.]
  {A. Ortiz-Gil$^{1,2}$,
  L. Guzzo$^2$, P. Schuecker$^3$, H. B\"ohringer$^3$, C.A. Collins$^4$\\
  $^1$Observatorio Astron\'omico, Universidad de Valencia, Aptdo. 22085, Valencia, E-46071-SPAIN\\
  $^2$INAF, Osservatorio Astronomico di Brera, Merate, I - 23807, Italy\\
  $^3$Max-Planck-Institut f\"ur Extraterr. Physik, D 85740 Garching, Germany\\
 $^4$Liverpool John Moores University, Liverpool, U.K.}
\date{Released 2003 Xxxxx XX}

\pagerange{\pageref{firstpage}--\pageref{lastpage}} \pubyear{2003}

\def\LaTeX{L\kern-.36em\raise.3ex\hbox{a}\kern-.15em
    T\kern-.1667em\lower.7ex\hbox{E}\kern-.125emX}

\begin{document}

\label{firstpage}

\maketitle

\begin{abstract}

We present an estimate of the bolometric X-ray luminosity - velocity dispersion ($L_x - \sigma_v$) 
relation measured from a new, large and homogeneous sample of 171 low redshift, X-ray selected galaxy 
clusters. The linear fitting of $\log(L_x) - \log(\sigma_v)$ gives 
$ L_x = 10^{32.72 \pm 0.08} \sigma^{4.1 \pm 0.3}_v$ erg s$^{-1}$ h$^{-2}_{50}$.
Furthermore, a study of 54 clusters, for which the X-ray temperature of the intracluster medium
T is available, allows us to explore two other scaling relations, $L_x -T$ and $\sigma_v -T$. From this sample
we obtain $L_x \propto T^{3.1 \pm 0.2}$ and $\sigma_v \propto T^{1.00 \pm 0.16}$, which are
fully consistent with the above result for the $L_x$-$\sigma_v$.
The slopes of $L_x -T$ and $\sigma_v -T$ are incompatible with the 
values predicted by self-similarity ($L_x \propto T^{2} \propto \sigma_v^4$), thus
suggesting the presence of non-gravitational energy sources heating up the 
intracluster medium, in addition to the gravitational collapse, in the early stages 
of cluster formation. On the other hand, the result on $\log(L_x) - \log(\sigma_v)$
supports the self-similar model. 

\end{abstract}

\begin{keywords}
 cosmology: observations -- galaxies: clusters: general -- galaxies: fundamental
parameters
\end{keywords}

\section{Introduction}

Galaxy clusters are dark matter haloes which form mainly through the
process of gravitational collapse, and so they are expected to be self-similar:
small clusters or galaxy groups must be scaled-down versions of
the more massive systems. Self-similarity predicts a series of relationships
between different cluster observables such as X-ray luminosity $L_x$, mass $M$,
temperature $T$ and galaxy velocity dispersion $\sigma_v$, which are the
same in all systems (Kaiser 1991).

The radial velocity dispersion 
$\sigma_v$ of the galaxies in a cluster probes the depth and shape of the potential well,
assuming that the luminous matter
traces reasonably well the dark matter in clusters.
Moreover, the intracluster gas emits X rays through a process of thermal
bremsstrahlung and its bolometric luminosity $L_x$ is found to be strongly
correlated with $\sigma_v$, as both gas and galaxies share a common potential
well (Solinger \& Tucker 1972). Also, since the X-ray emission can be well 
modelled by thermal emission of a hot, optically thin plasma, $L_x$ and the 
plasma temperature $T$ must be correlated. Finally, because galaxies are
embedded in the intracluster medium, a correlation is also expected between
the gas temperature $T$ and the galaxies velocity dispersion $\sigma_v$.

Self-similar models consider that the only energy source in the cluster
comes from the gravitational collapse, predicting the following scaling relations:
$L_x \propto T^{2} \propto \sigma_v^{4}$. Whereas there seems to be a general
agreement between different measurements that $L_x \propto T^{ \sim 3}$,
the measurement of $L_x - \sigma_v$ has given somewhat contradictory results
so far. Some authors have found that $L_x \propto \sigma^{4}$ indeed,
although with quite
large measurement errors or rather small data samples, while
others find slopes larger than 4 (see Table \ref{literature}).
Part of the differences in the results could come from different 
ways of selecting the sample, with a preference for regular clusters
in some of these surveys.

It has also been suggested that clusters and groups do not follow the same
$L_x- \sigma_v$ scaling relation, the latter being flatter than the
former (e.g. Mahdavi et al. 2000, Xue \& Wu 2000, 
Dell'Antonio et al. 1994). But there are other measurements contrary to 
that conclusion (Mahdavi \& Geller 2001, Mulchaey \& Zabludoff 1998).

For more distant clusters ($z$ between $ \sim 0.15$ and $\sim 0.6$) 
there is some evidence that the slope is also $> 4$ (Borgani et al. 1999,
Girardi \& Mezetti 2001), although only small samples are available 
at the moment, and more data are needed to reduce the error bars.

There is a clear need of new measurements performed on large samples (more
than 100 clusters), selected from a homogeneous dataset. The largest
cluster (not group) samples available up to date, on which this kind of 
study has been performed, are literature compilations combining data
from different authors and/or instruments (for example Xue \& Wu 2000, 
Mahdavi \& Geller 2001).
The present work comes to partially fill this gap, as we have a large 
homogeneous subsample of clusters selected from REFLEX, a flux-limited 
X-ray selected cluster catalogue built under well defined selection 
criteria. 

The structure of this paper is as follows. In Section \ref{thedata} we 
give a brief introduction to the REFLEX catalogue and the way in which the 
subsample in this paper has been selected. Section 
\ref{measurement} illustrates in detail the process of the $L_x$ and
$\sigma_v$ measurements, which lead to the the fitting 
of the $\log(L_x) - \log(\sigma_v)$ in Section \ref{lxsigma}. In
Section \ref{nosubst} we propose the removal of those clusters in
the sample which are multiple systems.
In Section \ref{discussion}
we analyse possible biases due to the nature of the selected sample. A study
on the $L_x - T$ and $\sigma_v - T$ relation follows in Section \ref{lxt},
and we finish with a summary in Section \ref{summary}. We are assuming 
H$_0=50$ km s$^{-1}$ Mpc$^{-1}$ in a flat universe with $\Omega _{\Lambda}=0.7$.

\section{The data}
\label{thedata}

The REFLEX (ROSAT-ESO Flux-Limited X-ray) cluster survey (B\"ohringer
et al 2001, B\"ohringer et al 2003 in preparation, Guzzo et al. 2003 
in preparation) has identified and measured successfully the redshift for all southern 
galaxy clusters (at galactic latitude $|b_{II}| > 20^o$) down to a flux limit 
$f_x > 3 \cdot 10^{-12}$ erg s$^{-1}$ cm$^{-2}$  in the ROSAT 
 All-Sky Survey. This sample has already provided us with 
results concerning the measurement of the cluster X-ray luminosity 
function at low redshift (B\"ohringer et al 2002), the power spectrum (Schuecker
et al. 2001a), the spatial correlation function (Collins et al. 2000),
the measurement of $\Omega_{\rm m}$ and $\sigma_8$ (Schuecker et al. 2003a) or
the measurement of the equation of state parameter $w$ of the dark energy 
(Schuecker et al. 2003b),
among others. A detailed description of the sample construction
can be found in B\"ohringer et al. (2001).

REFLEX is an X-ray selected
sample and the first compilation of the REFLEX catalogue includes 452 clusters 
down to a flux limit of 
$3 \times  10^{-12}$ erg s$^{-1}$ cm$^{-2}$ in the 
ROSAT energy band from 0.1 to 2.4 keV. 
The survey covers a total area in the sky of 4.24 ster. Two regions
corresponding to the Magellanic clouds and the Galactic plane have been
excluded because of the difficulties in identifying clusters of galaxies
in these densely crowded regions, as well as some strips of the sky which
were not observed by ROSAT due to power-off of the detector when the satellite
was crossing the Earth's radiation belts (see Figure 2 in B\"ohringer
et al 2001 for a survey exposure map). 

Optical follow-up has been performed through long-slit and multi-object
spectroscopy at the 1.5m, 2.2m and 3.6m ESO telescopes at La Silla (Chile) during
a series of observing campaigns, which started back in 1991. 
We complement our sample of galaxy redshifts by looking into the literature
(using the NASA/IPAC Extragalactic Database, NED) and adding those redshifts 
already published for the previously known clusters.

The cluster membership of each individual galaxy has been decided based
on the search for velocity peaks in the velocity distribution of the galaxies
observed in the field. All galaxies were assigned the same weight in the computation
of the cluster's median redshift because in most cases not enough photometric
material is available to give proper weights (i.e. to identify a real cD
galaxy). Anyway, Bright Cluster Galaxies (if present) were used to decide
on the main redshift of the system when it showed multiple velocity peaks in
the redshift histogram along the line of sight. In other cases the choice among
multiple velocity peaks was decided taking into account the correlation between
the peak position and the X-ray emission centre and also the
strength of the peak. Finally, other clusters ended up being
split into different systems when multiple X-ray emission centres 
associated to them were detected.

\begin{figure}
\begin{center}
\includegraphics[width=9.cm]{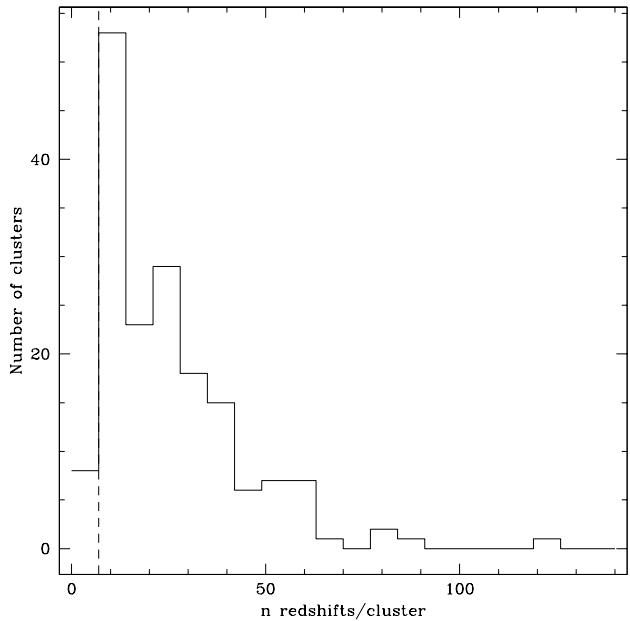}
\caption{Distribution of the number of redshifts per cluster in the 171 subsample of the
REFLEX  catalogue which we study in this paper.
The dashed vertical line shows the limit of 7 redshifts per cluster imposed to select the 
clusters for which a $\sigma_v$ is measured.}
\label{z_per_clu}
\end{center}
\end{figure}

\begin{figure}
\begin{center}
\includegraphics[width=8.cm]{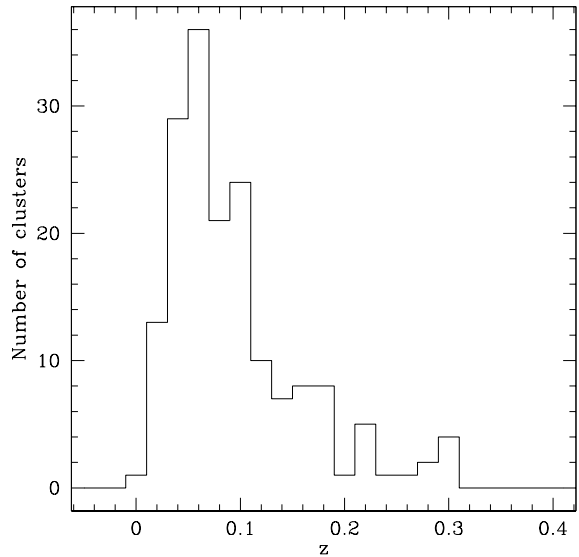}
\caption{Redshift distribution of the 171 clusters in the selected sample
from the REFLEX catalogue.}
\label{z_histo}
\end{center}
\end{figure}

We have selected from REFLEX those clusters with more than seven measured galaxy 
redshifts (see Fig. \ref{z_per_clu}), including only galaxies at a 
maximum projected distance from the cluster center of 0.5 h$_{50}^{-1}$ Mpc,
due to the lack of data at larger distances from the cluster centre.
Outliers in the velocity space were rejected by applying the 3 sigma clipping
method: galaxies with velocity differences with the cluster centre velocity larger than
three times the measured velocity dispersion were not used in the subsequent
iterations to avoid possible contamination by foreground/background objects. 
By following this procedure we end up with 171 clusters out of the 452 in the whole catalogue.
It is a low-redshift sample, the median redshift being 0.076.
Only 14 clusters have  $z$ above 0.2 (see Fig. \ref{z_histo}).

\section{Measurement of $L_x$ and $\sigma_v$.}
\label{measurement}

\subsection{Bolometric X-ray luminosity $L_x$}

Counts in the ROSAT images were measured using the
Growth Curve Analysis (GCA) and then converted into fluxes, as described
in B\"ohringer et al. (2001). Luminosities have
first been transformed into the ROSAT rest-frame 
band (0.1-2.4 keV) (see B\"ohringer et al. 2002 for details) and 
afterwards into bolometric luminosities.
We assumed a MEKAL spectral model with
an ICM metallicity of 0.3 Z$_\odot$ (Anders \& Grevesse 1989). We 
used XSPEC (Arnaud 1996) to obtain the relation between the luminosity in the 
0.1-2.4 keV band and the bolometric one, as a function of the cluster 
temperature(see B\"ohringer et al. 2003 for a detailed conversion table), 
which was computed from the velocity dispersions assuming
the empirical relation between $\sigma_v$ and T measured by
Girardi et al. (1996).

\subsection{Radial velocity dispersion $\sigma_v$}
\label{sigma}

The radial velocity dispersions $\sigma_v$ have been computed 
by means of the biweight estimator of central location and scale (within ROSTAT
by Beers, Flynn \& Gebhardt 1990). This estimator was first suggested by Tuckey 
(in Andrews et al. 1927) as a better way to study non-gaussian or contaminated normal
distributions than the gaussian estimators (mean and standard deviation).
The biweight location estimator belongs to the family of estimators known
as $M$ estimators of location. It works by minimizing a function of the deviations
of each observation from the estimate of location and
has the advantage that it does not assume an underlying gaussian
distribution, which in our case means to have a perfectly relaxed cluster
and might not be the case in most of the sample. It is less affected by 
points in the wings of the distribution
and is robust for a broad range of non-gaussian underlying populations.

We used the biweight estimator of central location and the gapper estimator 
of scale (based on the gaps between order statistics, Wainer \& Thissen, 1976)
when the number of redshifts available in the cluster was $7 \leq N_{z} \leq 10$. 
For $N_{z} > 10$ the biweight estimator was chosen for both 
location and scale. Errors were obtained in all cases by jacknifing of the
biweight. The choice of the different estimators has been done on the basis of
the tests carried out by Beers et al. (1990) for samples with different
number of data points. 
The usual cosmological correction and the correction for velocity errors 
(Danese, de Zotti \& Tullio, 1980) were also applied. 
The whole procedure was performed iteratively until the results converged.

The clusters
with the larger errors in the velocity dispersion estimate are among the
ones for which $N_{z} \leq 10$.

One source of concern is the fact that we are computing the $\sigma_v$ using
galaxies which are quite close to the cluster centre, at distances less than
0.5 h$^{-1}_{50}$ Mpc (median distance is 0.04$^{-1}_{50}$ Mpc). Girardi et al. 
(1996) find that the value of $\sigma_v$
is a function of the distance to the cluster centre, and it reaches a
 stable value when galaxies at a projected distance of 
$\sim $ 1 h$^{-1}_{50}$ Mpc are considered. At smaller distances, the  $\sigma_v$ 
does not show a constant behaviour: sometimes it raises up, sometimes it goes
down, and in some cases it remains stable. So, in a large
sample like ours we expect that this effect will cancel out.
From our current data, it is not possible to determine the $\sigma_v$ at
clustercentric distances larger than 0.5 h$^{-1}_{50}$ Mpc.

\section{The $L_x- \sigma_v$ relation}
\label{lxsigma}

\begin{figure}
\begin{center}
\includegraphics[width=9.cm]{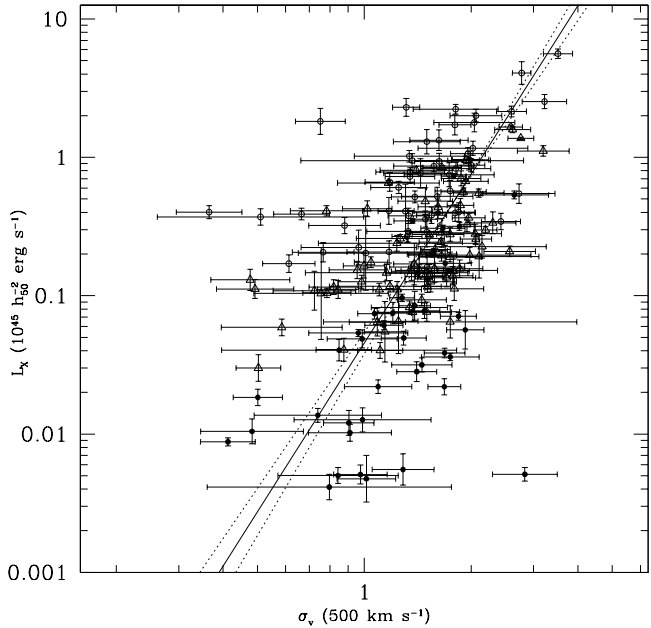}
\caption{Bolometric X-ray luminosity $L_x$ vs. radial velocity dispersion $\sigma_v$
for a REFLEX sample of 171 clusters. The linear fit gives a log-slope of $4.1 \pm 0.3$.
and the dashed lines are the 1-sigma errors. Filled circles correspond to clusters at redshift 
$z \leq 0.05$, open triangles are clusters with $0.05 < z \leq 0.1$ and open circles  are clusters 
at $ z > 0.1$.}
\label{odr_data_all}
\end{center}
\end{figure}

A power law was fit to the
$\log(L_x)- \log(\sigma_v)$ relation by means of the Orthogonal Distance Regression 
method (ODR, Boggs et al. 1992), which takes into account errors
on both variables. We took the data in units of $L_x / 10^{45}$ erg s$^{-1}$ ($L_{45}$) and
velocity dispersions in units of $\sigma_v / 500$ km s$^{-1}$ ( $\sigma_{500}$), that is, we put the origin of
coordinates more or less at the center of the datapoint cloud to help the model find the right
Y-axis intercept. We will use these units throughout the paper, but for Table \ref{literature},
where we quote $L_x$ in erg s$^{-1}$ and $\sigma_v$ in km s$^{-1}$ for comparison purposes
with the works referenced there.
The best fit to $ \log (L_{45}) - \log ( \sigma_{500}) $
is found to be (see Figure \ref{odr_data_all}) 

\begin{equation}
 \log(L_{45})= (-1.34 \pm 0.08)+ (4.1 \pm 0.3) \cdot \log ( \sigma_{500} )\ \rmn{erg} \ \rmn{s}^{-1} 
\rmn{h}^{-2}_{50}
\end{equation}

The Spearman rank correlation coefficient for this sample is $r_s=0.51$ and the
probability that a random distribution had this value of $r_s$ or larger 
by chance is $P \leq 2.6 \ 10^{-11}$. 

\begin{figure}
\begin{center}
\includegraphics[width=9.cm]{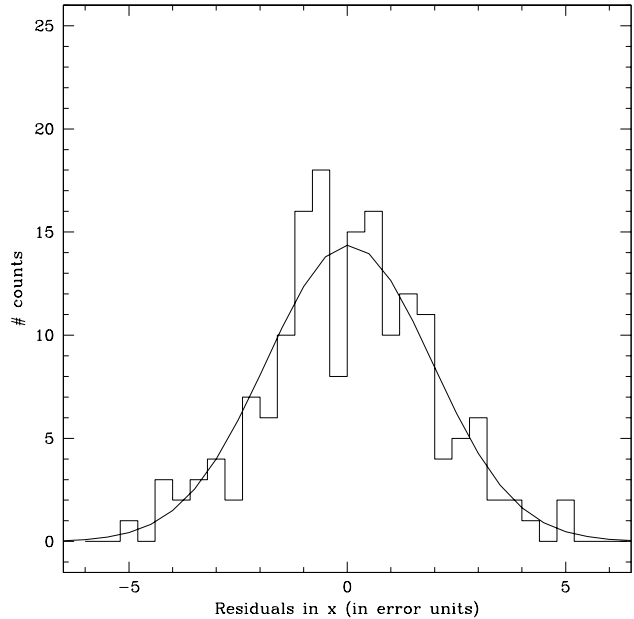}
\caption{Distribution of residuals along the X axis in the 171 cluster sample. We also show the
corresponding maximum likelihood normal distribution. The distribution of residuals
is well approximated by a gaussian distribution, with central location at $\bar{x} = 0.04$ 
and standard deviation $\sigma=1.9$.}
\label{gaussianity_x_7}
\end{center}
\end{figure}

The distribution of errors in the sample has been assumed to be
normal in the ODR fitting. To test for this assumption, we have performed a maximum likelihood 
analysis with sigma clipping to find which is the best fitting gaussian to 
the actual error distribution in X. As errors in X are far larger than errors in Y,
we can neglect the effect of the latter. We find that the best gaussian has central location
$\bar{x} = 0.04$ and standard deviation $\sigma=1.9$ (Figure \ref{gaussianity_x_7}).The error distribution along the X axis is thus reasonably gaussian.

We can also perform a crude estimate of the intrinsic dispersion of the sample,
$\Sigma_{int}$,
\begin{equation} \Sigma_{int}=\sqrt{ ( \sigma_{obs}^2 - \sigma_{meas}^2)} ,
\label{eqno2}
\end{equation}
in error units. $\sigma_{obs}$ is the best fitting gaussian $\sigma$ parameter as found
above for the current error distribution, that is $\sigma_{obs}=1.9$. $\sigma_{meas}$ 
is the $\sigma$ parameter of the error distribution in the ideal case in which only
the measurement errors are responsible for the dispersion in the data, which would be a 
gaussian with $\sigma=1$. We find $\Sigma_{int}=1.62 $ in error units, that is, the intrinsic 
dispersion of the sample is slightly larger than the dispersion due to the errors in 
the data. To see what this means in terms of physical units, let us assume that a typical value 
of the error in the velocity dispersion is the median of the error distribution,  
$\Sigma_{typ}=120$ km s$^{-1}$. The intrinsic dispersion is then 

\begin{equation} 
\sigma_{int}=1.62 \cdot \Sigma_{typ}=195 \  \rmn{km} \ \rmn{s}^{-1}
\end{equation}

Finally, to test for the stability of the fitting, we repeated it by removing those points
with an error in the velocity dispersion larger than 30 percent of their value. 
We find 
\begin{equation}
 \log(L_{45})= (-1.38 \pm 0.09)+ (4.2 \pm 0.3) \cdot \log(\sigma_{500}),
\end{equation} 
thus showing the robustness of the fitting against the presence of large errors in $\sigma_v$.

\begin{figure}
\begin{center}
\includegraphics[width=9.cm]{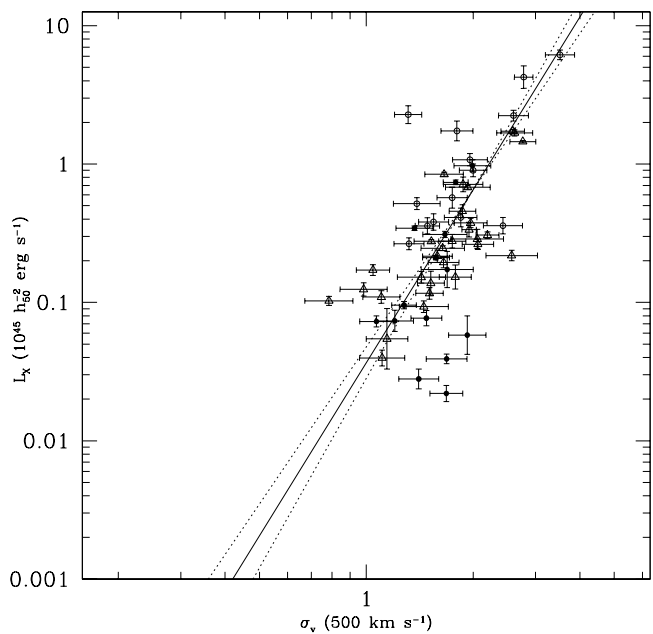}
\caption{Bolometric X-ray luminosity $L_x$ vs. radial velocity dispersion $\sigma_v$
for a REFLEX sample of 57 clusters for which more than 30 galaxy redshifts entered the
$\sigma_v$ computation. The linear fit gives a log-slope of $4.1 \pm 0.4$.
and the dashed lines are the 1-sigma errors. Filled circles correspond to clusters at redshift 
$z \leq 0.05$, open triangles are clusters with $0.05 < z \leq 0.1$ and open circles  are clusters 
at $ z > 0.1$.}
\label{odr_data_30}
\end{center}
\end{figure}

We may also worry about the small number of redshifts (7 at minimum) with which we are computing the $\sigma_v$ in some clusters. We have repeated the fit,this time taking only clusters for which $\sigma_v$ had been determined from at least 30 individual redshifts. In this way, we are left with a sample of only 57 clusters. The best fit relation is 
\begin{equation}
 \log(L_{45})= (-1.44 \pm 0.12)+ (4.2 \pm 0.4) \cdot \log(\sigma_{500}),
\end{equation} 
which is in agreement at the one sigma confidence level with the fit obtained for 
the whole sample. Figure \ref{odr_data_30}
shows the best fit found.

The best fitting gaussian to the error distribution in X is centered at 
$\bar{x} = 0.3$ and has standard deviation $\sigma=1.9$. This gives a value for the 
intrinsic dispersion of $\Sigma_{int}=195$ km s$^{-1}$.

\section{Discussion}
\label{discussion}

Self-similarity in the purely gravitational hierarchical
scenario of structure formation implies that for galaxy clusters 
$L_x \propto T^{2} \propto \sigma_v^{4}$. 
Although several authors have found a good agreement
between this prediction and their measurements for the $L_x- \sigma_v$
relation, others have found steeper values (see Table \ref{literature}). 
 
\begin{table*}
\caption{Compilation of literature values obtained for $a$ and $b$ in the 
relation $\log(L_x)=a+b \times \log(\sigma_v)$. The number of clusters that entered
the computation is also shown, together with some information about the sample
and the cluster selection method. $L_x$ is in units of erg s$^{-1}$ and $\sigma_v$ is in
km s$^{-1}$. All of them, except for Quintana \& Melnick 1982, make use of bolometric luminosites.
Only White et al. 1997 remove cooling flow clusters from the sample. Multicomponent clusters
are explicitely removed in Girardi \& Mezzetti 2001, Borgani et al. 1999 and in this work(b).This
work(c) is the result obtained when using a volume-limited sample.}

\label{literature}
\begin{tabular}{c | c | c | c | c | c}
\hline
Reference          & $b$            & $a$                 & No. clusters & Flux-limited sample & Selection method \\
\hline
\cite{edge}        & $ 2.90 \pm 0.19$    & $36.60 \pm 0.55$     & 23   & no          &  optical       \\
\cite{quintana}    & $3.7 \pm 0.4  $     &    -                 & 31   & no          &  optical               \\
\cite{mulchaey}    & $4.29 \pm 0.37$     & $31.61 \pm 1.09$     & 38   &    no  & optical \& X-ray         \\
\cite{mahdavi1}    & $4.4^{+0.7}_{-0.3}$ & $31.8^{+0.9}_{-2.0}$ & 280  &  no & optical \& X-ray       \\
\cite{girardi}     & $4.4^{+1.8}_{-1.0}$ & $29.4^{+3.0}_{-5.4}$ & 51   &  no & optical \& X-ray          \\
\cite{borgani}     & $5.1^{+1.2}_{-0.8}$ & $27.8^{+3.0}_{-2.2}$ & 53   &  yes        & X-ray       \\
\cite{xue}         & $ 5.30 \pm 0.21 $   & $28.32 \pm 0.61$     & 197  &  no & optical \& X-ray                \\
\cite{white}       & $5.36 \pm 0.16    $ & $39.3^{+0.13}_{-0.9}$ & 14   &  no & optical \\
this work(a)          & $4.1 \pm 0.3$       & $32.72 \pm 0.08$     & 171  &  yes        &  X-ray       \\
this work(b)          & $4.2 \pm 0.4$       & $32.41 \pm 0.10$     & 123   &  yes        &  X-ray       \\
this work(c)        & $  3.2 \pm 0.3$       & $35.16 \pm 0.09$     & 51    &  yes        & X-ray        \\
\hline
\end{tabular}
\end{table*}

Our value of $a=4.1 \pm 0.3$ is 
in good agreement with the slope measured by \cite{quintana},\cite{mulchaey}, \cite{mahdavi1} 
and \cite{girardi}. 
As for the intercepts, our result is compatible with 
\cite{mulchaey} and \cite{mahdavi1} at about the one-sigma confidence level.

\begin{figure}
\begin{center}
\includegraphics[width=8.cm]{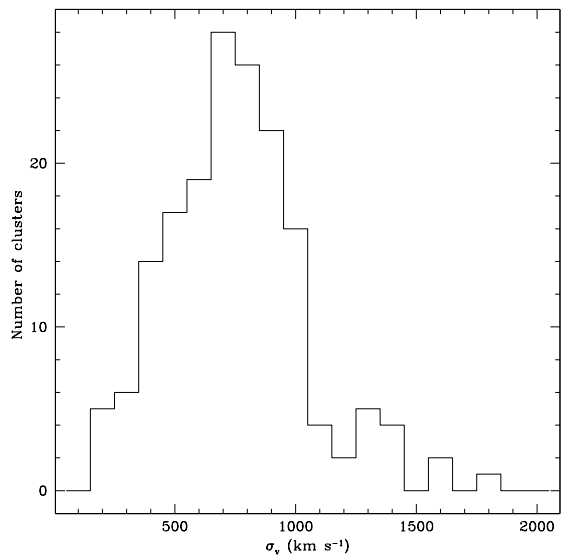}
\caption{Velocity dispersion distribution in the cluster sample}
\label{sigmas_histo}
\end{center}
\end{figure}

From Fig. \ref{sigmas_histo} one can see that our sample is mostly
populated by clusters and groups are practically absent, if we
consider as ``groups'' those systems with $\sigma_v <$ 340 km s$^{-1}$
(following the criterium used by Mahdavi et al. 2000). More specifically, 
only 11 out of the 171 systems are groups. 
Some authors (e.g. Helsdon \& Ponman 2000, Mahdavi et al. 2000, Xue \& Wu 2000, 
Dell'Antonio et al. 1994) have found that the scaling laws
are different for clusters and groups. In particular,
they find that clusters show a steeper slope in the
$L_x- \sigma_v$ relation.
Unfortunately, e cannot derive a meaningful $L_x- \sigma_v$ relation for groups from our 
sample, due to the small number of groups observed. 

The complexity involved in the measurements of $\sigma_v$ and $L_x$
(because of the presence of substructures, cooling flows, and other
physical phenomena), the largely unknown intrinsic dispersion of 
their relationship, and the use of different fitting methods could
explain the different results reached by the different works.
For example, \cite{xue} illustrate the different results obtained by using
ODR regression and Ordinary Least Squares (OLS) regression in the analysis.

\subsection{A sample free from clusters with substructures.}
\label{nosubst}

Substructures in clusters constitute a problem in the radial velocity
dispersion estimates, as the different components show different $\sigma_v$ and
an estimate of a single $\sigma_v$ for the cluster may not be meaningful.
We cannot make a reliable study of the substructures in our clusters as
in many cases we do not have enough redshifts to perform it. The sensitivity
for substructure detection depends strongly on the number of galaxies with
measured redshift available for each cluster. In our sample, though, we have 
large differences in the number of galaxies per cluster and trying to
detect substructure is not feasible. 

In a paper by Schuecker et al (2001b) they identify the clusters in the 
REFLEX sample which show multiple components in the X-ray emission. When
we take those clusters out from our sample, we end up with 123 clusters.
In this case we have

\begin{equation}
 \log(L_{45})= (-1.28 \pm 0.10)+ (4.2 \pm 0.4) \cdot \log(\sigma_{500}),
\end{equation} 

which is again in very good agreement with the one obtained for the whole
sample.

This result shows that  our sample is not largely affected
by the problems of substructure, probably because we are selecting galaxies 
very close to the X-ray peak emission, that is, to the central potential well, 
where one would expect that the strength of the gravitational potential will 
erase any substructures which might be present.
Even in the case of redshifts taken from the literature (see Section \ref{in_redshifts}) we
have not considered galaxies farther than 0.5 h$^{-1}_{50}$ Mpc, to be consistent with
the characteristics of our own data.





\subsection{Inhomogeneous redshift data sources}
\label{in_redshifts}

The galaxy redshifts used in this work come from three different sources.
When a cluster was already known and well studied, redshifts were taken
from the literature, possibly from different authors too. When a cluster
was already known but had little data available in the literature, we took
our own redshifts in order to secure and/or improve the cluster redshift.
Finally, in the case of clusters discovered in this project, a complete
optical follow-up has been carried out. 

It is thus not unreasonable to think that such a variety of redshift
sources might introduce some undesirable effect particularly on the $\sigma_v$ estimate
and therefore in the final $L_x - \sigma_v$ relation. To check whether this is the
case, we have restricted ourselves to a sample of 38 clusters for which
more than 7 redshifts per cluster were available, all of them being
redshifts measured by us from our own data. This is then
to be considered a completely  homogeneous sample in terms of both 
luminosity and redshift, as both quantities have been measured in
the same way for all 38 clusters. The result of the ODR fitting is 
$\log(L_{45})=(-0.63 \pm 0.10) + (2.2 \pm 0.4) \times \log(\sigma_{500})$.
In this case the fit is quite poor, probably because this sample
suffers from important selection effects. First, as we have now 
a small number of data points we should worry about the problems of
measuring $\sigma_v$ with galaxies at small projected distance from the cluster
centre, as explained in Section \ref{sigma}.

The minimum number of galaxy redshifts set to measure the  $\sigma_v$ may have a role too,
specially in this case where not too many clusters are available for
the fitting. When we consider the whole sample, clusters with smaller
number of individual galaxies are given much less weight into the
fit than those with a larger number of observed galaxies. That is not
the case here, as all of the clusters show small numbers (only 3 have
more than 15 measured $z$), so the differences in the weights are not 
so dramatic. 

\subsection{Is this a representative sample from the whole REFLEX catalogue?}

Of course, we would like to use all 452 clusters in the REFLEX 
catalogue to estimate the $L_x - \sigma_v$ relation. This cannot be done, 
unfortunately, as the survey was targeted to measure cluster redshifts and
not velocity dispersions in these clusters. While Multi-Object Spectroscopy 
observations were used for a few distant, compact clusters (due to the 
small field of view, 5 arcmin, of the EFOSC spectrograph on the ESO 3.6m 
telescope, see e.g. Guzzo et al. 1999), providing around 15 redshifts per cluster,
most of the new REFLEX cluster redshifts come from (multiple) single slit
observations which do not deliver enough galaxy redshifts to compute a velocity 
dispersion, although the main goal was always to have at least 5 spectra
per target.

\begin{figure}
\begin{center}
\includegraphics[width=8.cm]{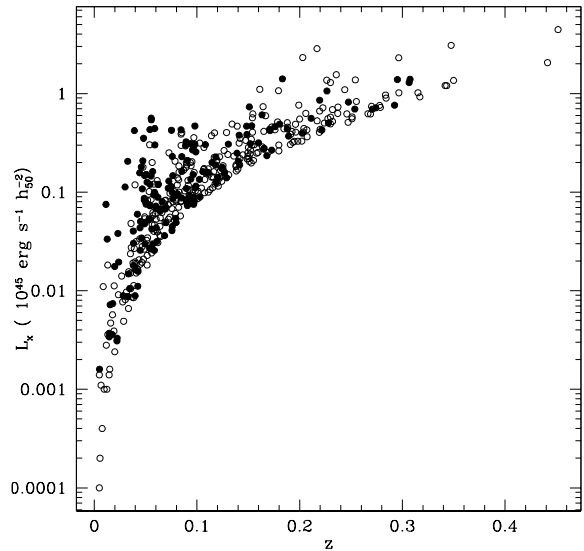}
\caption{The redshift - X-ray luminosity [0.1-2.4 keV] distribution for all clusters in
the REFLEX catalogue (open circles) and those from the 81 cluster subsample
used in this work (filled circles). }
\label{lx_z}
\end{center}
\end{figure}

We have compared the $L_x - z$ distribution for the whole catalogue
with the one corresponding to the 171 cluster subsample used in this work.
Figure \ref{lx_z} shows as filled circles the clusters from the subsample, and
as open circles the ones from the whole REFLEX catalogue. One can see
that both samples are approximately equally distributed. The results 
from a Kolmogorov-Smirnov
test are compatible with this conclusion: we find a maximum distance value
between the two samples (the entire catalogue and the 171 cluster sample)
of 0.11, the probability of this value being larger
if coming from a random sample being 0.22 .

\subsection{Bias from a flux-limited sample: A volume-limited subsample}
\label{vlted}

A bias is expected when measuring the $L_x - \sigma_v$ relation 
from a flux-limited sample. Specifically, at the flux cut-off level 
only the brightest clusters will be considered, for a given velocity 
dispersion. As our sample has actually
been drawn from a flux-limited sample, we may worry about whether we
are introducing such a bias towards larger luminosities. 

This bias can be avoided by considering a volume-limited subsample.
Therefore, we have repeated the analysis for the largest volume-limited
sample that we could construct from the 171 cluster sample.
It was built by taking all clusters with luminosity
$L_x  \geq 0.5 \times 10^{44}$ h$_{50}^{-2}$ erg s$^{-1}$, which
corresponds to a redshift limit of $z_{max}=0.08$. To this redshift
the comoving volume surveyed by REFLEX is $147.4 \cdot 10^{6}$ h$^{-3}_{50}$ Mpc$^3$

We have a total of 51 ``useful'' clusters (according to the selection
criteria outlined in Section \ref{thedata}) out of the 88 REFLEX clusters in 
this volume. The ODR fitting gives 
$\log(L_{45})= (-1.20 \pm 0.09) + (3.2 \pm 0.3) \times \log(\sigma_{500})$,
only compatible at the three-sigma error level with the one obtained from
the whole sample, which is an indication of some bias present in the
total sample.

\section{The $ L_x - T$ and $ \sigma_v - T$ relations}
\label{lxt}

Another important scaling relation in galaxy clusters is the X-ray
luminosity-temperature $ L_x - T$.
Again, measurements show that it does not follow the 
scaling relations
predicted by pure gravitational collapse models
($L_x \propto T^{2}$). The observed $L_x - T$ relation
is closer to $L_x \propto T^3$ (e.g. Novicki et al. 2002, Fairley et al. 2000, Xue \& Wu 2000,
Arnaud \& Evrard 1999, Reichart et al. 1999), which seems to indicate the presence of 
additional non-gravitational heating sources in the ICM. 

$\sigma_v - T$ has also been studied in the literature. There seems to be
a consensus in that $\sigma_v \propto T^{\approx 0.6}$ (see for example Xue \& Wu 2000, 
Girardi et al. 1996, Bird, Mushotzky \& Metzler 1995, Lubin \& Bahcall 1993), again
not in agreement with what would be expected in a pure gravitational collapse 
scenario, $\sigma_v \propto T^{0.5}$.

\subsection{$ L_x - T$ and $ \sigma_v - T$ in the entire sample}
We have studied these relations within our cluster sample to see to which degree
our data confirm or otherwise those results.
ASCA temperatures from two different catalogues (Horner 2001 and Ikebe et al. 2002)
were used for 54 clusters included in our  171 cluster sample. Ikebe et al.'s measurements
were preferred whenever both works had data for the same cluster, as they used a 
two-temperature model (instead of Horner's one component model) for the isothermal 
plasma, allowing a multiphase intra-cluster medium, thus obtaining more accurate T estimates. 

\begin{figure}
\begin{center}
\includegraphics[width=8.cm]{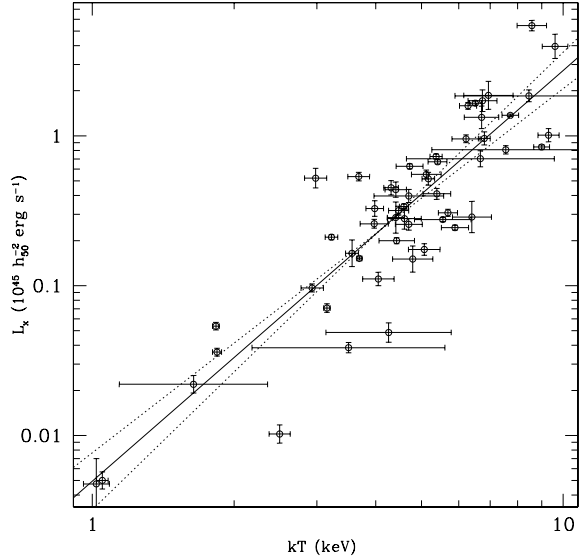}
\caption{$L_x - T$ for a subsample of 54  
clusters drawn from the total sample, for which ASCA temperatures were available.
The solid line is the ODR best-fit model, with one-sigma errors marked as dashed 
lines.}
\label{lx_t}
\end{center}
\end{figure}

\begin{figure}
\begin{center}
\includegraphics[width=8.cm]{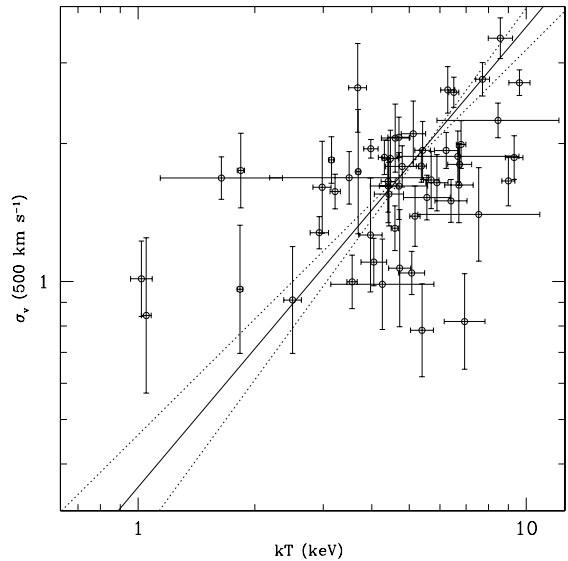}
\caption{$\sigma_v - T$ for a subsample of 54  
clusters drawn from the total sample, for which ASCA temperatures were available.
The solid line is the ODR best-fit model, with one-sigma errors marked as dashed 
lines.}
\label{vdisp_t}
\end{center}
\end{figure}

Again, we performed the ODR fitting in the log-log space, obtaining the following best-fit
relations (see also Figures \ref{lx_t} and \ref{vdisp_t}) :

\begin{equation}
\log(L_{45})= (-2.53 \pm 0.16) + (3.1 \pm 0.2) \cdot \log(T)
\end{equation}
\begin{equation}
\log(\sigma_{500})= (-0.45 \pm 0.11) + (1.00 \pm 0.16) \cdot \log(T)
\end{equation}

The $L_x - T$ estimate  that we find is in good agreement (at one-sigma confidence
level) with the 
slope and intercepts previously measured by Fairley et al. 2000 and 
Arnaud \& Evrard 1999. 

For the $\sigma_v - T$ relation, we are finding here a larger slope than
found by other authors (which is $\approx 0.6$). To check that we are not 
biased for including clusters 
with substructure in the sample, we select a new set of clusters from the
sample in Section \ref{nosubst} for which temperature data are available.
We have 33 systems, for which we find

\begin{equation}
\log(\sigma_{500})= (-0.38 \pm 0.13) + (0.92 \pm 0.18) \cdot \log(T)
\end{equation}

in agreement with the previous result.
Furthemore, we can select only clusters with the most reliable $\sigma_v$ determination,
that is, those with $\sigma_v$ measured from more than 30 galaxy redshifts. In doing
so, we end up with a sample of 16 clusters, giving

\begin{equation}
\log(\sigma_{500})= (-0.37 \pm 0.15) + (0.9 \pm 0.2) \cdot \log(T)
\end{equation}

\subsection{$ L_x - T$ and $ \sigma_v - T$ in the voulme limited sample}

It is interesting to probe these relations also in a volume-limited sample.
From the one we constructed in Sect. \ref{vlted}, we select those clusters
for which temperature data are available, ending up with a 28 cluster sample.
On this we find that

\begin{equation}
\log(L_{45})= (-2.7 \pm 0.3) + (3.2 \pm 0.4) \cdot \log(T)
\end{equation}
\begin{equation}
\log(\sigma_{500})= (0.49 \pm 0.05) + (0.77 \pm 0.19) \cdot \log(T)
\end{equation}

The $L_x -T $ is still the same, but the $\sigma_v - T $ is flatter than
what is found in the 54 cluster sample, steeper than what is expected
from self-similar models (at one-sigma confidence level), and in good
agreement with what is found by other authors (Xue \& Wu 2000, Girardi et al. 1996
and Lubin \& Bahcall 1993, for example).

In conclusion, we have derived an $L_x - T$ similar to what is found by previous
studies, and which is in contradiction with self-similarity. But the 
$\sigma_v - T $ we are measuring is not reliable, as it has shown to
vary depending on the sample's selection criteria. The scatter and
the large error bars in the velocity dispersions make it impossible to
obtain a consistent relation. Still, we can draw a useful conclusion, 
which is that self-similarity is ruled out in any case.

\section{Summary and conclusions}
\label{summary}

We present an estimate of the $L_x - \sigma_v $ relation 
computed with a new large and homogeneous sample of clusters from the 
REFLEX survey. All studies on the $L_x - \sigma_v$ relation in 
large samples (with more than $\approx 100$ clusters) have been performed so far on 
compilations from the literature requiring the use of models to combine different data
from different authors and/or instruments (e.g. Xue \& Wu 2000, 
Mahdavi \& Geller 2001). 

From the REFLEX catalogue we construct a sample of 171 clusters for
which we can derive reliable estimates of $\sigma_v$, and then fit the 
$ \log(L_x)-\log(\sigma_v) $ relation, finding

\begin{equation}
 \log(L_{45})= (-1.34 \pm 0.08)+ (4.1 \pm 0.3) \cdot \log ( \sigma_{500})\ \rmn{erg} \ \rmn{s}^{-1} 
\rmn{h}^{-2}_{50}
\end{equation}

in agreement with what is expected from self-similar models of cluster
formation.

The large size and homogeneity of the sample allows us to make an estimate 
of the intrinsic dispersion of the $L_x - \sigma_v$ relation. We find that 
$\Sigma_{int}= 195$ km s$^{-1}$.

Measurements of $L_x$ and $\sigma_v$ in clusters with multiple components 
may not be reliable. The identification in the literature 
of those clusters allows us to build a new sample where we
remove those with significant substructure in their X-ray emission 
in Schuecker et al. (2001b). This procedure
leaves us with a sample of 123 clusters, on which we fit again the 
$ \log(L_x)-\log(\sigma_v) $ relation, finding

\begin{equation}
 \log(L_{45})= (-1.28 \pm 0.10)+ (4.2 \pm 0.4) \cdot \log ( \sigma_{500} )\ \rmn{erg} \ \rmn{s}^{-1} 
\rmn{h}^{-2}_{50}
\end{equation}

\noindent
compatible at one-sigma confidence level with the slope 
and intercept that we find when using the 171 cluster sample,
suggesting that the sample is not noticeably affected by clusters
with multiple components. The fact that we are considering only galaxies
close to the cluster centre may be a reason for that.

We have also investigated the $L_x- \sigma_v $ relation in a volume-limited 
sample, finding $\log(L_{45})= (-1.20 \pm 0.09) + (3.2 \pm 0.3) \times \log(\sigma_{500})$.
This significantly flatter slope is an indication that some bias may be present on 
the whole sample due to the flux limit imposed on the data.

The $L_x - T$ and $\sigma_v - T $ scaling relations have also been explored,
finding that $L_x \propto T^{3.1 \pm 0.2}$ and $\sigma_v \propto T^{1.00 \pm 0.16}$,
in a subsample of 54 clusters for which $T$ was available from the literature.
These results are consistent (within the error bars) with the one obtained for the 
$L_x - \sigma_v$, as we find $ L_x \propto \sigma_v^{3.1 \pm 0.7}$ when using 
these $L_x - T$ and $\sigma_v - T$ to derive $L_x - \sigma_v$.

The slope found in the $ L_x - T $ relationship is steeper 
than the value predicted by a purely gravitational collapse model of 
cluster formation, and is in good agreement 
with previous measurements performed on other cluster samples. 
As for the $ \sigma_v - T $, the slope we find is only compatible with
the self-similar value at the three-sigma confidence level, and slightly 
larger than the value found by other authors. Also, if we consider a
volume-limited sample, the slope found is then in good agreement with 
previous studies, yet incompatible with self-similarity.

The fact that this model fails to reproduce the measured slope is attributed
to the contribution to the ICM energy budget from other non-gravitational 
heating sources (see e.g. Bialek, Evrard \& Mohr 2002), like AGNs, 
supernovae, gas cooling (Pearce et al. 2000), the presence of cool cores 
(Allen \& Fabian 1998), or 
the effect of shocks in cluster formation (Cavaliere et al. 1997).

Here we are faced with a paradoxical result: the slope of $L_x - \sigma$ 
supports the self-similar scenario, whereas $ L_x - T $ and $\sigma_v - T$ 
do not, and all three relations are consistent with each other within one sigma
confidence level. Clearly, more data are required to clarify the situation
and reduce the error bars.
This is a reflection of the general situation in this topic in the literature:
so far there is a general agrement on $ L_x - T $ and $ \sigma_v - T $, but 
there is not such an agreement on the $L_x - \sigma_v$.

\section{Acknowledgements}

We thank C. Altucci as her laurea thesis served as a starting point
and guideline for this work. We are grateful to the whole REFLEX
team which helped in the compilation of this catalogue. We are
also indebted to Mrs. Yu Xing, who provided us with the data
needed to compute the bolometric luminosity out of the luminosity
in the ROSAT band.
We also thank M. Girardi and S. Borgani
for stimulating discussions and suggestions. 
We are indebted to T. Beers who gently provided us with the 
ROSTAT routines, as well as to the authors of the ODR package, P. Boggs
and J. Rogers. PS acknowledges support by  DLR under the grant No.\,50\,OR\,9708\,35.
This research has made use of the NASA/IPAC Extragalactic Database (NED) 
which is operated by the Jet Propulsion Laboratory, California Institute 
of Technology, under contract with the National Aeronautics and Space Administration. 
This research has made use of NASA's Astrophysics Data System.

\end{document}